\documentclass[longauth]{aa}  

\usepackage{graphicx}
\usepackage{txfonts}
\usepackage{hyperref}
\usepackage{natbib}
\usepackage{float}
\usepackage{color}
\usepackage{enumitem}
\usepackage{amsmath}
\usepackage{amssymb}
\usepackage{orcidlink}

\bibpunct{(}{)}{;}{a}{}{,}

\newcommand{\ctref}{c_{t,\rm ref}}
\newcommand{\ctad}{c_{t,\rm ad}}
\newcommand{\ctfit}{c_{t,\rm fit}}

\begin{document} 

   \title{The ratio of horizontal to vertical displacement in solar oscillations estimated from combined SO/PHI and SDO/HMI observations}

\titlerunning{Horizontal to vertical displacement from SO/PHI and SDO/HMI}

   \author{J.~Schou\inst{1}\orcidlink{0000-0002-2391-6156}\thanks{\hbox{Corresponding author: J.~Schou} \hbox{\email{schou@mps.mpg.de}}} \and
   J.~Hirzberger\inst{1} \and
   D.~Orozco~Su\' arez\inst{2}\orcidlink{0000-0001-8829-1938} \and
   K.~Albert\inst{1}\orcidlink{0000-0002-3776-9548} \and
   N.~Albelo~Jorge\inst{1} \and
   T.~Appourchaux\inst{3}\orcidlink{0000-0002-1790-1951} \and 
   A.~Alvarez-Herrero\inst{4}\orcidlink{0000-0001-9228-3412} \and
   J.~Blanco Rodr\'\i guez\inst{5}\orcidlink{0000-0002-2055-441X} \and
   A.~Gandorfer\inst{1}\orcidlink{0000-0002-9972-9840} \and
   D.~Germerott\inst{1} \and
   L.~Guerrero\inst{1} \and
   P.~Gutierrez-Marques\inst{1}\orcidlink{0000-0003-2797-0392} \and
   F.~Kahil\inst{1}\orcidlink{0000-0002-4796-9527} \and
   M.~Kolleck\inst{1} \and
   S.K.~Solanki\inst{1}\orcidlink{0000-0002-3418-8449} \and
   J.C.~del~Toro~Iniesta\inst{2}\orcidlink{0000-0002-3387-026X} \and
   R.~Volkmer\inst{6} \and
   J.~Woch\inst{1}\orcidlink{0000-0001-5833-3738} \and 
   B.~Fiethe\inst{7}\orcidlink{0000-0002-7915-6723} \and
   I.~P\' erez-Grande\inst{9}\orcidlink{0000-0002-7145-2835} \and 
   E.~Sanchis~Kilders\inst{5}\orcidlink{0000-0002-4208-3575} \and
   M.~Balaguer~Jiménez\inst{2}\orcidlink{0000-0003-4738-7727} \and
   L.R.~Bellot~Rubio\inst{2}\orcidlink{0000-0001-8669-8857} \and
   D.~Calchetti\inst{1}\orcidlink{0000-0003-2755-5295} \and
   M.~Carmona\inst{8}\orcidlink{0000-0001-8019-2476} \and
   W.~Deutsch\inst{1} \and
   A.~Feller\inst{1} \and
   G.~Fernandez-Rico\inst{1,9}\orcidlink{0000-0002-4792-1144} \and
   A.~Fern\' andez-Medina\inst{4}\orcidlink{0000-0002-1232-4315} \and
   P.~Garc\'\i a~Parejo\inst{4}\orcidlink{0000-0003-1556-9411} \and 
   J.L.~Gasent~Blesa\inst{5}\orcidlink{0000-0002-1225-4177} \and 
   L.~Gizon\inst{1,10}\orcidlink{0000-0001-7696-8665} \and
   B.~Grauf\inst{1} \and 
   K.~Heerlein\inst{1} \and
   A.~Korpi-Lagg\inst{1}\orcidlink{0000-0003-1459-7074} \and
   A.~L\' opez Jim\' enez\inst{2} \and 
   T.~Maue\inst{6,11} \and 
   R.~Meller\inst{1} \and
   A.~Moreno Vacas\inst{2}\orcidlink{0000-0002-7336-0926} \and
   R.~M\" uller\inst{1} \and
   E.~Nakai\inst{6} \and 
   W.~Schmidt\inst{6} \and
   J.~Sinjan \inst{1}\orcidlink{0000-0002-5387-636X} \and
   J.~Staub\inst{1}\orcidlink{0000-0001-9358-5834} \and
   H.~Strecker \inst{2}\orcidlink{0000-0003-1483-4535} \and 
   I.~Torralbo\inst{9}\orcidlink{0000-0001-9272-6439} \and 
   G.~Valori\inst{1}\orcidlink{0000-0001-7809-0067} 
   }

   \institute{
         Max-Planck-Institut f\"ur Sonnensystemforschung, Justus-von-Liebig-Weg 3,
         37077 G\"ottingen, Germany
         \and
         Instituto de Astrofísica de Andalucía (IAA-CSIC), Apartado de Correos 3004,
         E-18080 Granada, Spain
         \and
         Univ. Paris-Sud, Institut d’Astrophysique Spatiale, UMR 8617,
         CNRS, B\^ atiment 121, 91405 Orsay Cedex, France
         \and
         Instituto Nacional de T\' ecnica Aeroespacial, Carretera de
         Ajalvir, km 4, E-28850 Torrej\' on de Ardoz, Spain
         \and
         Universitat de Val\`encia, Catedr\'atico Jos\'e Beltr\'an 2, E-46980 Paterna-Valencia, Spain
         \and
         Leibniz-Institut für Sonnenphysik, Sch\" oneckstr. 6, D-79104 Freiburg, Germany
         \and
         Institut f\"ur Datentechnik und Kommunikationsnetze der TU
         Braunschweig, Hans-Sommer-Str. 66, 38106 Braunschweig,
         Germany
         \and
         University of Barcelona, Department of Electronics, Carrer de Mart\'\i\ i Franqu\`es, 1 - 11, 08028 Barcelona, Spain
         \and
         Instituto Universitario "Ignacio da Riva", Universidad Polit\'ecnica de Madrid, IDR/UPM, Plaza Cardenal Cisneros 3, E-28040 Madrid, Spain
         \and
         Institut f\"ur Astrophysik, Georg-August-Universit\"at G\"ottingen, Friedrich-Hund-Platz 1, 37077 G\"ottingen, Germany
         \and
         Fraunhofer Institute for High-Speed Dynamics, Ernst-Mach-Institut, EMI, Ernst-Zermelo-Str. 4, 79104 Freiburg, Germany
}

\date{Received January 19, 2023; accepted March 15, 2023}

 
\abstract{
In order to make accurate inferences about the solar interior using helioseismology, it is essential to understand all the relevant physical effects on the observations.
One effect to understand is the (complex-valued) ratio of the horizontal to vertical displacement of the p- and f-modes at the height at which they are observed. 
Unfortunately, it is impossible to measure this ratio directly from a single vantage point, and it has been difficult to disentangle observationally from other effects.
In this paper we attempt to measure the ratio directly using 7.5~hours of simultaneous observations from the Polarimetric and Helioseismic Imager on board  Solar Orbiter and the Helioseismic and Magnetic Imager on board the Solar Dynamics Observatory.
While image geometry problems make it difficult to determine the exact ratio, it appears to agree well with that expected from adiabatic oscillations in a standard solar model. On the other hand it does not agree with a commonly used approximation, indicating that this approximation should not be used in helioseismic analyses. 
In addition, the ratio appears to be real-valued.
}


\maketitle
%
%

\section{Introduction}
As the duration of helioseismic observations has increased, systematic errors have become a major concern.
In global helioseismology this has been studied in great detail by, for example, 
\cite{2004ApJ...602..481K} and \cite{2015SoPh..290.3221L}, and in local helioseismology by \cite{2012ApJ...749L...5Z} and \cite{2018A&A...619A..99L}.
\cite{2015SoPh..290.3221L} showed that one of the main effects that
needs to be taken into account is that the displacement caused by the oscillations is not purely vertical.

Given the importance of knowing the horizontal to vertical ratio, several attempts have been made to estimate it, including those
by \cite{1998ApJ...504L.131S}, \cite{1998ESASP.418..933K}, \cite{1999A&A...346..633S}, \cite{2001ApJ...561.1127R}, and \cite{2004ApJ...602..481K}.
Unfortunately, 
none of these investigations used
direct measurements of the different velocity components as they are difficult to obtain from a single vantage point, where a Doppler measurement will only give one of the three components. In principle, the transverse components could be obtained from correlation tracking, but this is technically difficult and would involve combining velocities from different methods, which will in turn likely have different effective observing heights.
Instead, the estimates were made using indirect methods. \cite{1998ApJ...504L.131S} used a ring diagram analysis and analyzed the azimuthal dependence of the power, while others compared the magnitude of leaks in a global mode analysis with those from theoretical models.
Unfortunately, these indirect methods all depend on assumptions about physical and instrumental effects such as the height dependence of the observations and the instrumental point spread function, leaving significant uncertainties. 
Having said that, none of these measurements clearly indicated substantial deviations from the theoretical expectations (see the next section).

Solar Orbiter \citep[SO;][]{2020A&A...642A...1M} was launched in 2020, 
with the Polarimetric and Helioseismic Imager (SO/PHI) instrument \citep{2020A&A...642A..11S} on board.
A key feature of SO is that it spends most of the time away from the Sun-Earth line.
In addition, SO/PHI and the Helioseismic and Magnetic Imager (HMI) instrument \citep{2012SoPh..275..229S} on board the Solar Dynamics Observatory \citep[SDO;][]{2012SoPh..275....3P}
observe the same spectral line.
It is thus now possible to obtain measurements of the oscillations from two different vantage points, observed in a consistent manner, which is exploited here to estimate the horizontal to vertical displacement ratio.

In Sect.~\ref{Theory} we discuss the theory, in Sect.~\ref{Observations}
the data used, in Sect.~\ref{Method} the analysis method, and in Sect.~\ref{Results} some results.
Finally we discuss the results in Sect.~\ref{Discussion} and present our conclusions in Sect.~\ref{Conclusion}.

\section{Theory}
\label{Theory}

Solar oscillations cause both vertical and horizontal motions near the solar surface. 
The vertical motion of an undamped monochromatic wave with unit vertical amplitude is given by
\begin{equation}
\label{Vr}
V_r = \cos (k x - \omega t + \phi_0 )
\end{equation}
and the horizontal motion in the direction of propagation by
\begin{equation}
\label{Vh}
V_h = - c_t \sin (k x - \omega t + \phi_0 - \phi_h ),
\end{equation}
where a plane parallel geometry is assumed. Here $c_t$ is the ratio of the horizontal to vertical motion, $k$ the wave number, $x$ the horizontal distance in the direction of propagation, $\omega$ the angular frequency, $t$ the time, $\phi_0$ the initial phase of the mode, and $\phi_h$ the phase offset.
 We note that the sign convention was chosen so as to make $c_t$ positive for a surface gravity wave.
In the horizontal direction perpendicular to the direction of propagation, there is no transverse motion for the  nonmagnetic waves considered here.

Assuming adiabatic oscillations and a free surface, it can be shown \citep[see, e.g.,][]{2010aste.book.....A} that $c_t$ is given by
\begin{equation}
\ctref = \frac{g k}{\omega^2},
\label{eq:ctref}
\end{equation}
where $g$ is the surface gravity and $\phi_h = 0$.
Given that $c_t$ decreases strongly with frequency, the plots of other estimates of $c_t$ will be divided by this reference value to make any deviations more visible.
In this simple case the two motions are $90^\circ$ out of phase. 
For surface gravity waves (f-modes), where $c_t$ is close to unity, the two components are of equal magnitude resulting in a circular motion. 
For p-modes the motion is elliptical and becomes almost purely vertical at high frequencies.

Another way to estimate $c_t$ is to evaluate it numerically from adiabatic eigenfunctions calculated from a standard solar model. In the following, such estimates are denoted as   $\ctad$.

Observing from an angle $\alpha$ away from the surface normal and from a direction $\Delta\theta$ away from the direction of propagation, the observed velocity is
\begin{eqnarray}
\label{V1}
V &= & V_r \cos\alpha + V_h \sin\alpha\cos\Delta\theta \nonumber\\
&=& \cos (k x - \omega t + \phi_0 ) \cos\alpha \nonumber\\
&-& c_t \sin (k x - \omega t + \phi_0 - \phi_h) \sin\alpha\cos\Delta\theta .
\end{eqnarray}
This can, instead, be written as a sine wave with a different amplitude and phase:
\begin{eqnarray}
\label{V2}
V &=& A \cos (k x - \omega t + \phi_0 + \phi^\prime) \\
&=& A \cos (k x - \omega t + \phi_0 ) \cos \phi^\prime \nonumber
- A \sin (k x - \omega t + \phi_0 ) \sin \phi^\prime .\nonumber
\end{eqnarray}
Here $A$ is the amplitude and $\phi^\prime$ is the phase shift relative to $V_r$, with
\begin{eqnarray}
\label{phip}
\tan\phi^\prime
&=& \frac{ c_t \cos\phi_h\cos\Delta\theta\sin\alpha}
{\cos\alpha + c_t \sin\phi_h\cos\Delta\theta\sin\alpha}\\
&=& \frac{ c_t \cos\Delta\theta\tan\alpha\cos\phi_h}
{1 + c_t \cos\Delta\theta\tan\alpha\sin\phi_h}
.
\end{eqnarray}
If the phase is as expected ($\phi_h = 0$), then
\begin{equation}
\label{phip_simple}
\tan\phi^\prime  = c_t \cos\Delta\theta\tan\alpha.
\end{equation}
As expected from geometric considerations, observations from a vertical direction ($\alpha = 0^\circ$) should not show a phase difference with $\Delta\theta$;   for surface gravity waves, where $c_t=1$, the phase should equal the viewing angle when observing in the direction of propagation ($\Delta\theta=0$).
For small $c_t \tan\alpha$, the phase is proportional to $\cos\Delta\theta$, which as discussed in Sect. \ref{sec:fitted} becomes important.

The amplitude $A$ is also affected.
Unfortunately the observed amplitudes must be corrected for 
the velocity sensitivity of the two instruments, and in particular for their respective point spread functions.
As these values are not known to the required precision, no attempt is made to use the amplitude here.

\section{Observations and generation of observables }
\label{Observations}

For the analysis described here we used an eight-hour dataset taken by SO/PHI on March 23, 2021.
At this time SO was separated from SDO by about $108^\circ$ at a distance from the Sun of roughly 0.70~au.
These data were taken using the Full Disk Telescope (FDT) of SO/PHI with a cadence of roughly one dataset per 60~s, with each dataset containing a continuum image and five images taken across the 
\ion{Fe}{I}~6173~\AA\ line at nominal positions of -140~m\AA, -70~m\AA, 0~m\AA, 70~m\AA, and 140~m\AA\ relative to the center of the line.
All data were taken in a single linear (to the achievable accuracy) polarization state, which 
allows   a high cadence and minimizes the telemetry.
The images have a solar diameter of $\approx 770$ pixels. 
At the beginning of the observing period a number of images were also taken at offset tuning positions, but they were not used in the present analysis.
For HMI standard 45~s Dopplergrams from the JSOC dataseries hmi.V\_45s were used.

For the SO/PHI observations used here, the individual images were downlinked, which made it possible to optimize the algorithm used for calculating the Doppler velocity.
Specifically we chose to adapt the algorithm used by HMI, thereby making the data easy to combine.
This algorithm is based on the algorithm used for the Michelson Doppler Imager (MDI) and has thus become known as the MDI-like algorithm
\citep[see, e.g.,][]{2012SoPh..278..217C}.
In this algorithm two sums
\begin{equation}
\label{eq:cos}
C=\sum_{i=-2}^2 I_i \cos (2\pi i/5)
\end{equation}
and
\begin{equation}
\label{eq:sin}
S=\sum_{i=-2}^2 I_i \sin (2\pi i/5)
\end{equation}
are used to calculate a phase
\begin{equation}
\label{eq:phidop}
\phi_{\rm dop} = {\rm atan2}(S,C),
\end{equation}
where the $I_i$ represent the observed (dark corrected and flat-fielded) signals in the five images across the line, indexed with $i = 0$ corresponding to the 0~m\AA\ position, and a two-valued arctan (atan2) has been used.
From these the Doppler velocity can be estimated as
\begin{equation}
V = F(\phi_{\rm dop}),
\end{equation}
where $F$ is a lookup function.
To estimate $F$, a radiative transfer calculation was performed on a snapshot of a MURaM simulation of the near surface layers of the Sun. 
Specifically the Stokes-I profile from the G2 nonmagnetic case for the 6173~\AA\ line at $0^\circ$ viewing angle shown in Fig.~1 of \cite{2018A&A...617A.111S} was used.
These line profiles were simply averaged horizontally as the granulation is unresolved. 
This resulting line is then Doppler shifted, convolved with a Gaussian with a full width at half maximum (FWHM) of 0.0955~\AA\ to represent the filter profiles of the FDT, sampled at the tuning positions and passed through Eqs.~\ref{eq:cos}, \ref{eq:sin}, and \ref{eq:phidop}, from which $F$ is determined by interpolating the input Doppler shift as a function of calculated phase.

It should be noted that Eq.~\ref{eq:phidop} is insensitive to a wavelength-independent flat field. Multiplying all the data points with a constant   changes both $S$ and $C$ by the same factor, leaving $\phi_{\rm dop}$ unchanged. As the sum of the coefficients in Eqs.~\ref{eq:cos} and \ref{eq:sin} are both zero, it also follows that the derived velocity is independent of a wavelength-independent dark level.

\begin{figure}
\begin{center}
\includegraphics[width=0.95\columnwidth]{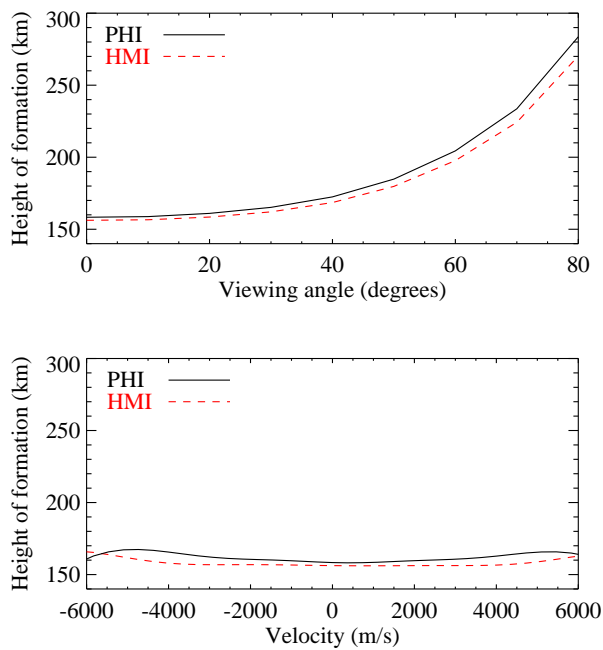}
\end{center}
\caption[]{
Effective observing height for the two instruments. Top: Height as a function of viewing angle at $V = 0$~m/s. Bottom: Height as a function of an added velocity at $0^\circ$ viewing angle.
The heights are calculated as described in the main text and are relative to the continuum $\tau=1$. 
For consistency, both plots have the same height range.
}
\label{fig:height}
\end{figure}

By adding a vertical velocity varying linearly with height to the simulation, it is possible to estimate the effective height of observation of the Doppler shifts as a function of viewing angle and Doppler velocity. Figure \ref{fig:height} shows these results, and those obtained for HMI using six tuning positions and the corresponding transmission profiles. The difference in the observing height between SO/PHI and HMI is only a few kilometers, which (as discussed in Sect.~\ref{sec:fitted}) turns out to be important. That the difference is so small is not very surprising. While the SO/PHI filters are somewhat wider than the HMI filters, the same line is used and the separation in wavelength is almost identical.

Once the Doppler velocity has been calculated at each pixel, the resulting Dopplergram is corrected for an estimate of the spatially dependent center wavelength caused by nonuniformities in the etalon used for the wavelength tuning. These corrected Dopplergrams are then undistorted based on an optical model of the instrument.
It should be noted that an error in this correction (or the effect of solar oblateness) is of little consequence for the present analysis. In the resulting tracked data (see Sect.~\ref{Method}) it appears, to lowest order,   as a pattern moving across the images at a rate given by the tracking velocity $v$, which in the resulting power spectrum   results in power at a frequency of $v k$, which is far below the frequencies of interest here, considering the range of wavenumbers used in the analysis.

To make use of the Dopplergrams it is also necessary to have the relevant metadata describing the image geometry and orbital parameters. In most cases these were taken directly from the keywords provided in the input data, but in a few cases they had to be recalculated, partly because of the preliminary nature of some of the current keywords. As the Dopplergrams are undistorted, the geometry keywords are no longer valid. To remedy this, a limb finding algorithm was run on an undistorted version of the continuum tuned filtergram and the relevant keywords were updated. The observing time for each dataset is given, in the input data, as the average of the time of the six filtergrams. For the purpose of measuring the Doppler shift, the continuum image does not contribute, and so the observing time is replaced by the average time of the five filtergrams taken across the line. Finally, the Carrington elements for the regular keywords were calculated without taking into account the Sun-SO light travel time. This results in an error in the Carrington longitude, and is thus corrected.

\section{Methods}
\label{Method}

\subsection{Combination of SO/PHI and HMI data}
\begin{figure}
\begin{center}
\includegraphics[width=0.95\columnwidth]{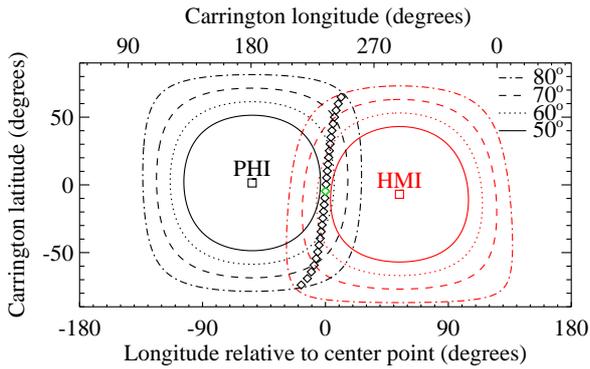}
\end{center}
\caption[]{
Observing geometry. The location of the SO/PHI and HMI sub-spacecraft points at the midpoint of the observations are shown by squares. The closed curves indicate the viewing angles  for the two instruments (see legend, top right corner) relative to the local surface normal. 
Diamonds indicate the locations used for the analysis.
The green diamond indicates the midpoint between the two instruments. The black diamonds are spaced by $5^\circ$.
}
\label{fig:geometry}
\end{figure}

To combine the SO/PHI and HMI data, both were tracked to the same grid in Carrington longitude and latitude.
In order to limit the data size, the HMI Dopplergrams (which have a solar diameter of about 3820 pixels versus about 770 pixels for SO/PHI) were first convolved with a Gaussian with FWHM of 4.0 input pixels and subsampled by a factor of two.
To remain consistent with the standard HMI ring diagram analysis, this is done using mtrack \citep{2011JPhCS.271a2008B} with a Postel mapping, producing $192 \times 192$ pixel tiles covering $15.36^\circ \times 15.36^\circ$ with a resolution of $0.08^\circ$ in each direction (in the following referred to as remapped pixels), as done in the standard HMI data analysis \citep{2011JPhCS.271a2009B}. The standard version of mtrack assumes 1.00~au when correcting for the finite distance, but the version here properly uses the SO distance of 0.70~au. To limit the effects of the solar rotation the data are tracked at the photospheric Doppler rate of \cite{1984SoPh...94...13S} at the center of each tile. The tile centers are located at  the set of longitudes and latitudes at which the viewing angles are the same from the two instruments at the center of the observing interval. Doing so ensures that any phase changes of the waves with height, such as those proposed to explain the center-to-limb systematics in helioseismology \citep{2012ApJ...760L...1B}, cancel out.
One tile is located at the midpoint of the two sub-spacecraft points and the rest are spaced by $5^\circ$ away from this (resulting in a significant overlap between adjacent tiles).
The tracking locations, as well as the location of the spacecraft in heliographic coordinates, are shown in Fig. \ref{fig:geometry}.

At each location the SO/PHI tiles are linearly interpolated to the HMI times (and excluding a single bad image), that is from a somewhat uneven 60~s cadence to a uniform 45~s cadence. The times used for the interpolation are those at which the light would have crossed 1~au. In the case of HMI these times are given by the keyword T\_REC, which by construction is uniformly spaced in solar time. For SO/PHI the time is calculated from the observing time and the distance to the Sun. 
To avoid the times when the SO/PHI tuning was offset from the spectral line (see Sect.~\ref{Observations}) and to have a number of times with low prime factors, only HMI T\_REC times from 2021.03.23\_04:39:45\_TAI through 2021.03.23\_12:09:00\_TAI, for a total of 600 HMI images or 7.5 hours, are used.
The temporal means for each remapped pixel and the spatial means at each time are then subtracted and the datacubes are apodized using the standard ring-diagram apodization \citep{2011JPhCS.271a2008B}.

In order to estimate the phase shifts, a Fourier transform is applied to the datacubes in time and space and the cross-spectra are calculated.
The resulting cubes are then, at each frequency, interpolated from $(k_x,k_y)$ to $(k,\theta)$, where $k_x = k \cos \theta$ is the wavenumber in the $x$ (longitude) direction, $k_y = k\sin \theta$ the wavenumber in $y$ (latitude) direction, $k = (k_x^2+k_y^2)^{1/2}$ is the total wavenumber, and $\theta$ the azimuth.  The grid in $k$ oversamples the original grid by a factor of 2.0. The grid in $\theta$ is chosen to also achieve a factor of 2.0 at the maximum $k$ used. These spectra are then 
interpolated to the mode frequencies for radial orders $n=0$ through 6, rebinned to 16 bins in $\phi$, and the phase is calculated.
To obtain the frequencies, the ring diagram fits from the Stanford JSOC dataseries hmi.rdvfitsf\_fd15 \citep{2011JPhCS.271a2009B} were used.
Specifically all valid fitted frequencies for Carrington rotation 2242 (which straddles the observing time for the present dataset) with a latitude of $0^\circ$ and longitude of $\pm 52.5^\circ$ (to most closely match the center location used here)
were averaged in time, at each $n$ and spherical harmonic degree $l=k R_\odot$, with $R_\odot$ being the solar radius. Finally these averaged frequencies were interpolated from $l$ to the $k$ values used here.

\subsection{Fitted model}
\label{sec:fitted}
To fit the observed data, the directions from the center of each tile to the two spacecraft have to be calculated: the angles to normal, $\alpha_{\rm PHI}$ and $\alpha_{\rm HMI}$, and the azimuths,
$\theta_{\rm PHI}$ and $\theta_{\rm HMI}$. From these quantities the azimuths needed for Eq.~\ref{phip} can be calculated as $\Delta\theta_{\rm PHI} = \theta-\theta_{\rm PHI}$ and $\Delta\theta_{\rm HMI} = \theta-\theta_{\rm HMI}$, and in turn  the expected phase difference can be obtained from Eq.~\ref{phip} as
\begin{equation}
\label{eqfit0}
\Delta\phi^\prime  = \phi^\prime_{\rm PHI} - \phi^\prime_{\rm HMI},
\end{equation} 
which can then be fitted to the data to estimate $c_t$.

\begin{figure}
\begin{center}
\includegraphics[width=0.95\columnwidth]{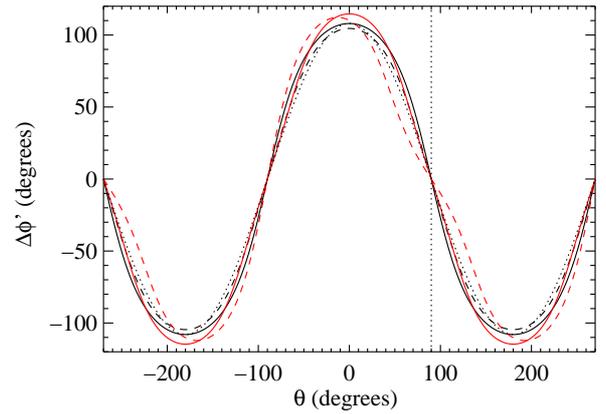}
\end{center}
\caption[]{
Examples of predicted phase differences between HMI and SO/PHI for an f-mode ($c_t = 1.0$) and $\alpha=54^\circ$.
The zero point of $\theta$ is chosen as the direction between the instruments.
Black curves show results for $\phi_h = 0$, red for $\phi_h = 45^\circ$.
Solid curves are for SO/PHI and HMI observing from opposite azimuths, corresponding to the green diamond in Fig. \ref{fig:geometry}.
Dashed curves are for a case with $140^\circ$ difference between the viewing azimuths, corresponding roughly to the points $30^\circ$ (six points) away from the green diamond in Fig. \ref{fig:geometry}.
The dotted black curve is a cosine, for reference.
The vertical dotted line shows the midpoint between the viewing directions, where the phase difference is zero.
The azimuth is shown over more than $360^\circ$ for clarity.
}
\label{fig:phip}
\end{figure}

Examples of the predicted phase differences are shown in Fig. \ref{fig:phip}.
Except when $\phi_h$ is nonzero and the viewing angles are not close to $180^\circ$ apart, the behavior is close to sinusoidal.

Unfortunately, it turns out that there are uncertainties in the image geometry, relative timing of the instruments, and the exact sensitivity with height in the atmosphere.
Locally an image geometry error, whether caused by errors in the geometry metadata, a height error or residual distortion will, to lowest order, result in a small spatial shift, $(\Delta x, \Delta y)$, which   in turn will cause an apparent phase shift of
\begin{equation}
\phi_{\rm geom}^\prime = k_x \Delta x + k_y \Delta y = k\Delta x \cos \theta + k\Delta y \sin \theta.
\end{equation}
Given the close to sinusoidal form of the expected phase difference, this and the geometric effect will be close to degenerate. Specifically, the geometric shift in the direction between the instruments will be close to degenerate ($\cos\theta$ in Fig.~\ref{fig:phip}), while the transverse term ($\sin\theta$ in Fig.~\ref{fig:phip}) will be close to orthogonal.

Similarly, a time offset $\Delta t$ between the two instruments will result in an apparent phase change of $\Delta\phi_{\rm time}^\prime = \omega\Delta t$, independent of $k_x$ and $k_y$.
For the simple case given by Eq.~\ref{phip_simple} the mean (over azimuth) of the expected phase shift is zero, and thus a constant offset will not affect the estimated $c_t$.
In the general case given by Eq.~\ref{phip} the mean is not zero. However, even in that case the mean phase difference between the two instruments is zero if $\alpha_{\rm PHI} = \alpha_{\rm HMI}$, as considered here.

To accommodate these sources of errors, while keeping the fit stable, the following equation is used:
\begin{equation}
\label{eqfit}
\Delta\phi^\prime = \phi^\prime_{\rm PHI} - \phi^\prime_{\rm HMI} + 
\Delta\phi_{\rm time}^\prime + k\Delta x_{\rm perp} \cos (\theta-\theta_{\rm mid}).
\end{equation}
Here $\theta_{\rm mid} = (\theta_{\rm PHI} + \theta_{\rm HMI})/2$ is the midpoint between the azimuths to the instruments and $\Delta x_{\rm perp}$ is the offset in the perpendicular direction.
At each $(k,n)$ the phases from the averaged and interpolated cross-spectra  are then fitted to obtain estimates of $c_t$ (in the following denoted $\ctfit$), $\Delta\phi_{\rm time}^\prime$, and $\Delta x_{\rm perp}$.

\section{Results}
\label{Results}
\begin{figure}
\begin{center}
\includegraphics[width=0.94\columnwidth]{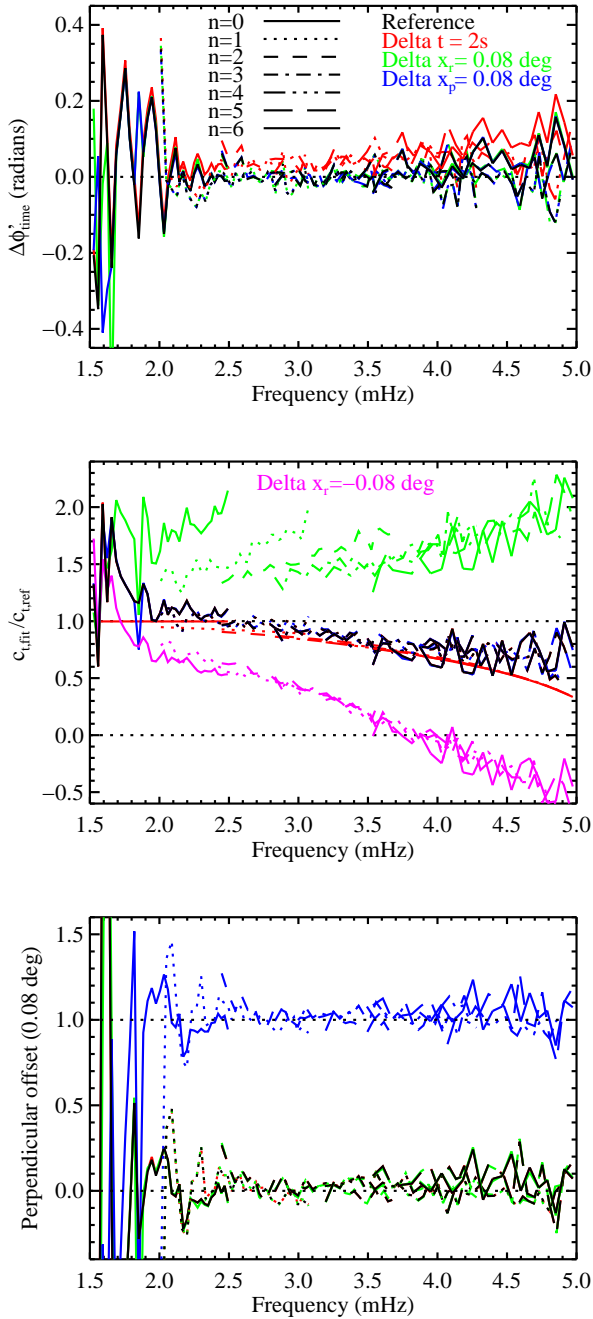}
\end{center}
\caption[]{
Fitted parameters for the target halfway between the sub-spacecraft points (green diamond in Fig.~\ref{fig:geometry}) assuming $\phi_h = 0$.
Results for each radial order $n$ are connected by lines (see legend in top panel).
In the middle panel the fitted $c_t$ values are divided by $\ctref$ for ease of display.
Black lines indicate the results with the nominal time shift and image offsets.
Red lines show the fits resulting from an arbitrary additional shift of 2~s between the instruments.
Green and blue lines show the results obtained by shifting the SO/PHI data relative to the HMI data by an additional 1.0~remapped pixel in the fitted cubes ($0.08^\circ$). The green lines use a shift in the direction of the maximum phase difference, the blue in the orthogonal direction.
In the middle panel the magenta lines show the results for a shift in the opposite direction (i.e., by -1.0 pixels vs +1.0 pixels for the green lines).
The black lines are plotted last, and hence the other colors are often invisible.
To avoid poor fits, only modes with $l \ge 200$, $k \le$ 0.9 times the Nyquist frequency for SO/PHI ($l \approx 620$) and $1.5~$mHz~$\le \nu \le 5.0$~mHz are shown.
The smooth red curves show $\ctad$.
}
\label{fig:results0}
\end{figure}

Figure \ref{fig:results0} shows the results for the target halfway between the sub-spacecraft points (green diamond in Fig.~\ref{fig:geometry}) assuming $\phi_h = 0$. 
For this analysis, offsets in time (2.3~s), $x$ (0.4 remapped pixels in the tracked tiles), and $y$ (-0.1 remapped pixels) were selected to make the results roughly agree with the expected values. 
Indeed, $\ctfit$ for the reference case (black lines) agrees fairly well with $\ctad$, and   the perpendicular shift and the time shift are both close to zero.

As discussed in the previous section, errors in the timing or geometry can be difficult to distinguish from the physical effect, as illustrated by the colored lines in Fig.~\ref{fig:results0}.
As expected, an artificial time shift only changes $\Delta\phi_{\rm time}^\prime$ (top panel) significantly. The change is also very close to the expected $\Delta\phi_{\rm time}^\prime = \omega\Delta t $.

An image offset in the direction of the expected maximum will change $\ctfit$ (middle panel of Fig.~\ref{fig:results0}) by an amount that depends on frequency and order. 
To judge the agreement with theory, the values of $\ctad$ are also shown, as estimated from eigenfunctions of Model S of \cite{1996Sci...272.1286C}. 
To match the observations, $\ctad$ was
evaluated at a height of 190~km above 
$\tau=1$, corresponding roughly to the effective height of the observations.
It should be noted that the $\ctad$ values agree well with those expected from an isothermal atmosphere, in particular with Eq.~(11) of \cite{1999A&A...346..633S}, for suitably chosen atmospheric parameters.
Attempting to fit for the shift together with $\ctfit$ does not work. The fits become unstable and the errors increase dramatically.

A shift in the orthogonal direction only changes the fitted offset $\Delta x_{\rm perp}$ in that direction (bottom panel of Fig.~\ref{fig:results0}).
The imposed shift of one remapped pixels changes the fitted offset by almost exactly one remapped pixel, as expected.
The fact that the inferred shifts appear to be independent of $n$ and frequency (and thereby of the degree and wavenumber) is consistent with a simple shift.  
Some optical aberrations such as coma
could have resulted in a $k$-dependent shift.

\begin{figure}
\begin{center}
\includegraphics[width=0.95\columnwidth]{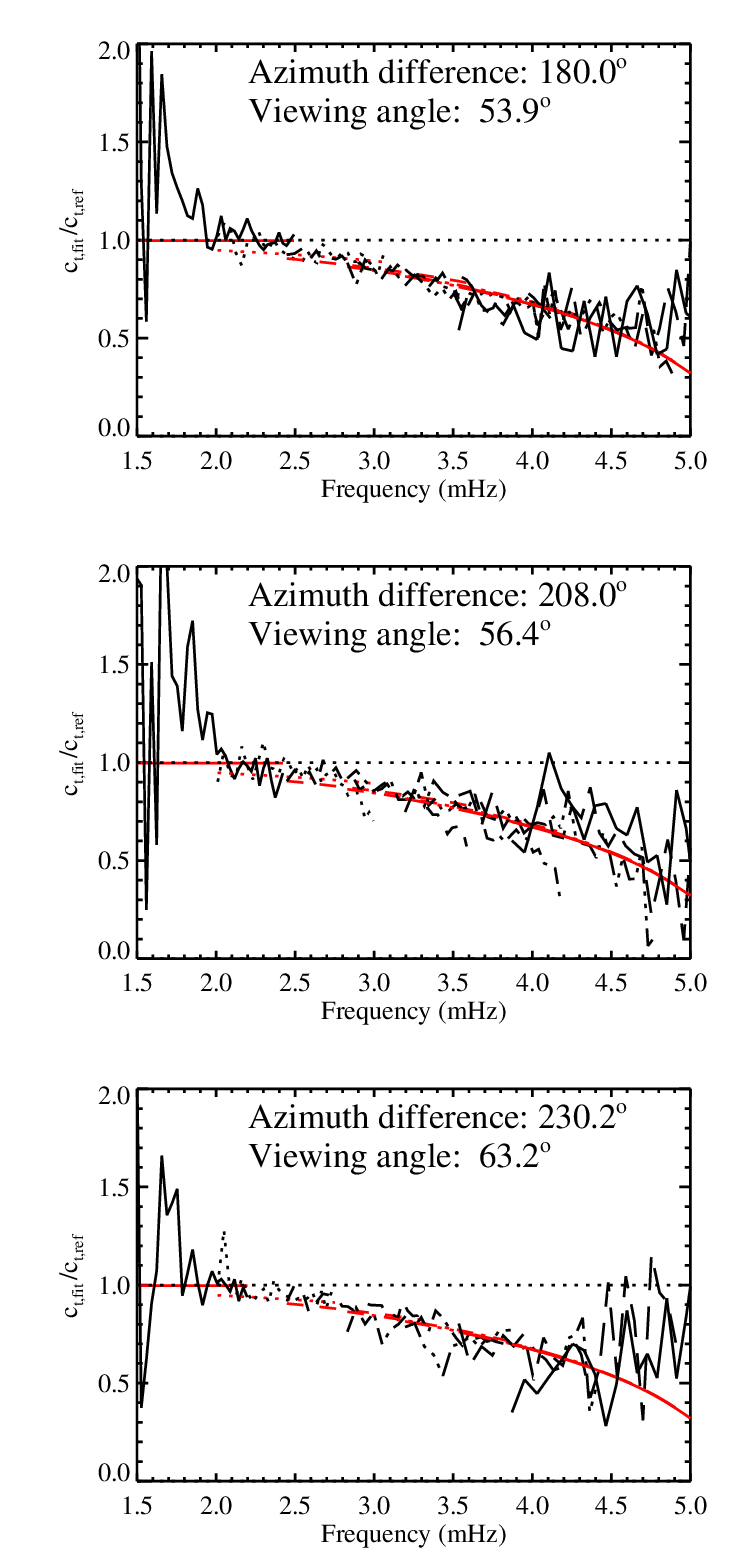}
\end{center}
\caption[]{
Ratios $\ctfit/\ctref$ for selected locations. The top panel corresponds to the green diamond in Fig.~\ref{fig:geometry}, the middle panel  to the black diamond $20^\circ$ above, and  the bottom panel to $40^\circ$ above.
To constrain the unknown shift, the choice where the mean ratio is 1.0 for 2.5~mHz~$\le \nu \le$~3.5~mHz was used.
In each panel the azimuth difference between SO/PHI and HMI is given, together with the viewing angle, which is common between the instruments.
As in Fig.~\ref{fig:results0}, the smooth red curves show $\ctad$ for a height of 190~km, corresponding to the top case. 
The $\ctad$ values for the heights corresponding to the two other cases are indistinguishable at the scale of the plots.
}
\label{fig:ctlat}
\end{figure}

The fact that the parameters change in an almost perfectly independent manner means that we are able to concentrate on the $c_t$ determination without having to worry about the other parameters.
It also means that one only needs to scan in one parameter to see how $\ctfit$ varies, and that a simple interpolation can be used to determine the results for arbitrary shifts. In particular,   a criterion for estimating the unknown shift can be selected and   the results interpolated to that. 
One possibility is to require 
that $\ctfit/\ctad = 1.0$ 
averaged over some frequencies. Results obtained by averaging over the modes with
 2.5~mHz~$\le \nu \le$~3.5~mHz are shown in Fig.~\ref{fig:ctlat} for selected viewing angles.
Clearly there is not much variation with viewing angle within the range shown,
and the variation with frequency agrees quite well with that of $\ctad$ over most of the frequency range. At very low and high frequencies there are some deviations, but they are likely due to the poor signal-to-noise ratio.
We note that since the image shifts were chosen to make the average correct around 3~mHz, the agreement there is not significant.

\begin{figure}
\begin{center}
\includegraphics[width=0.95\columnwidth]{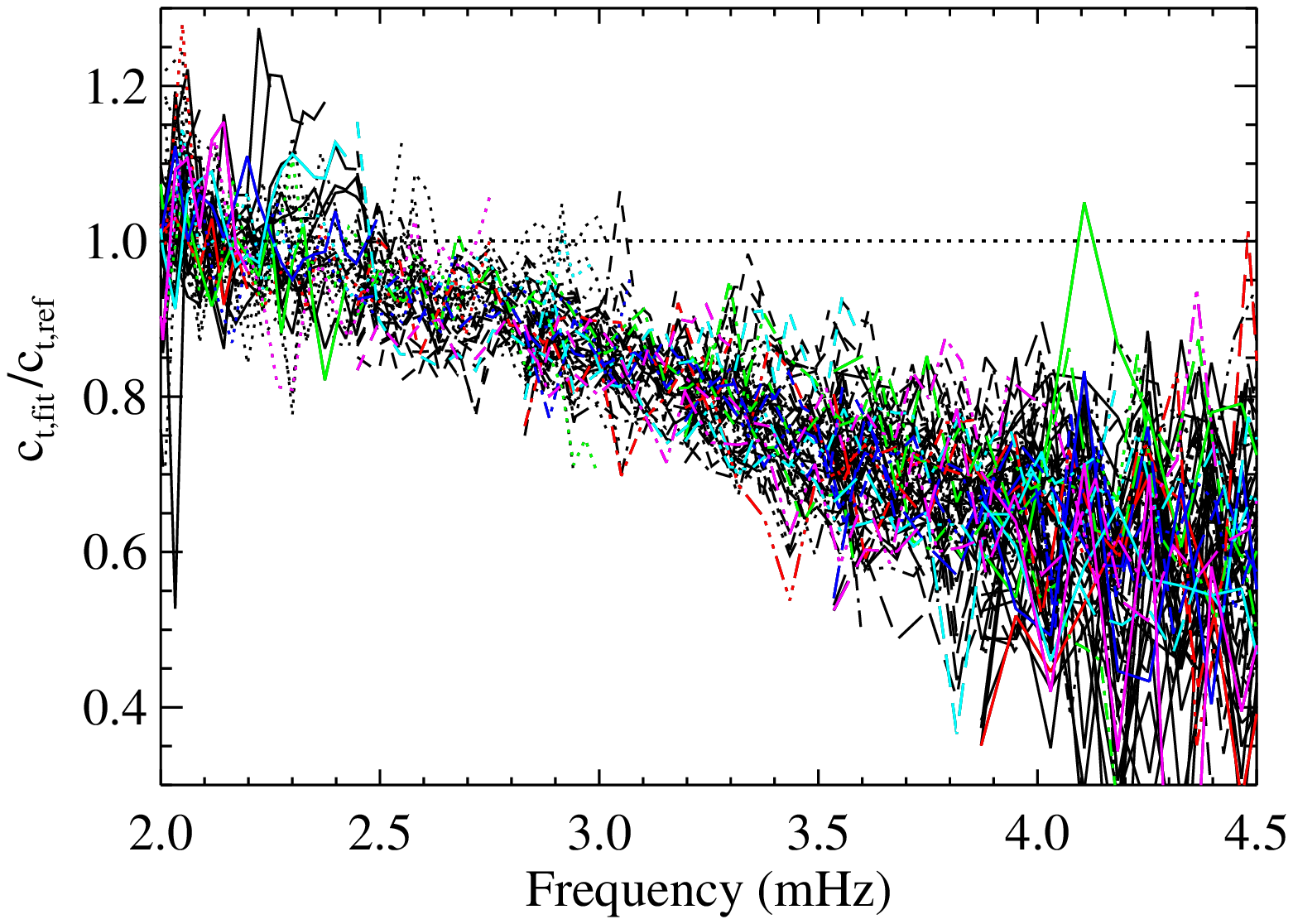}
\end{center}
\caption[]{
Same as  Fig.~\ref{fig:ctlat}, but plotted for all latitudes with a viewing angle of less than $70^\circ$.
Overplotted in color are (relative to the center position) $+40^\circ$ (red), $+20^\circ$ (green), $0^\circ$ (blue, corresponding to the green diamond), $-20^\circ$ (cyan), and $-40^\circ$ (magenta).
The plot is zoomed in relative to Fig.~\ref{fig:ctlat}.
}
\label{fig:ctlat1}
\end{figure}

To further illustrate this, Fig.~\ref{fig:ctlat1} shows the results for all the locations with a viewing angle of less than $70^\circ$ ($\pm$10 points in Fig.~\ref{fig:geometry}). While there is significant scatter, there are no obvious trends.
However, it is important to keep in mind that the averages were forced to agree for
2.5~mHz~$\le \nu \le$~3.5~mHz and that the set of fitted modes is latitude dependent. Nonetheless, the near uniform frequency dependence is striking.

\begin{figure}
\begin{center}
\includegraphics[width=0.95\columnwidth]{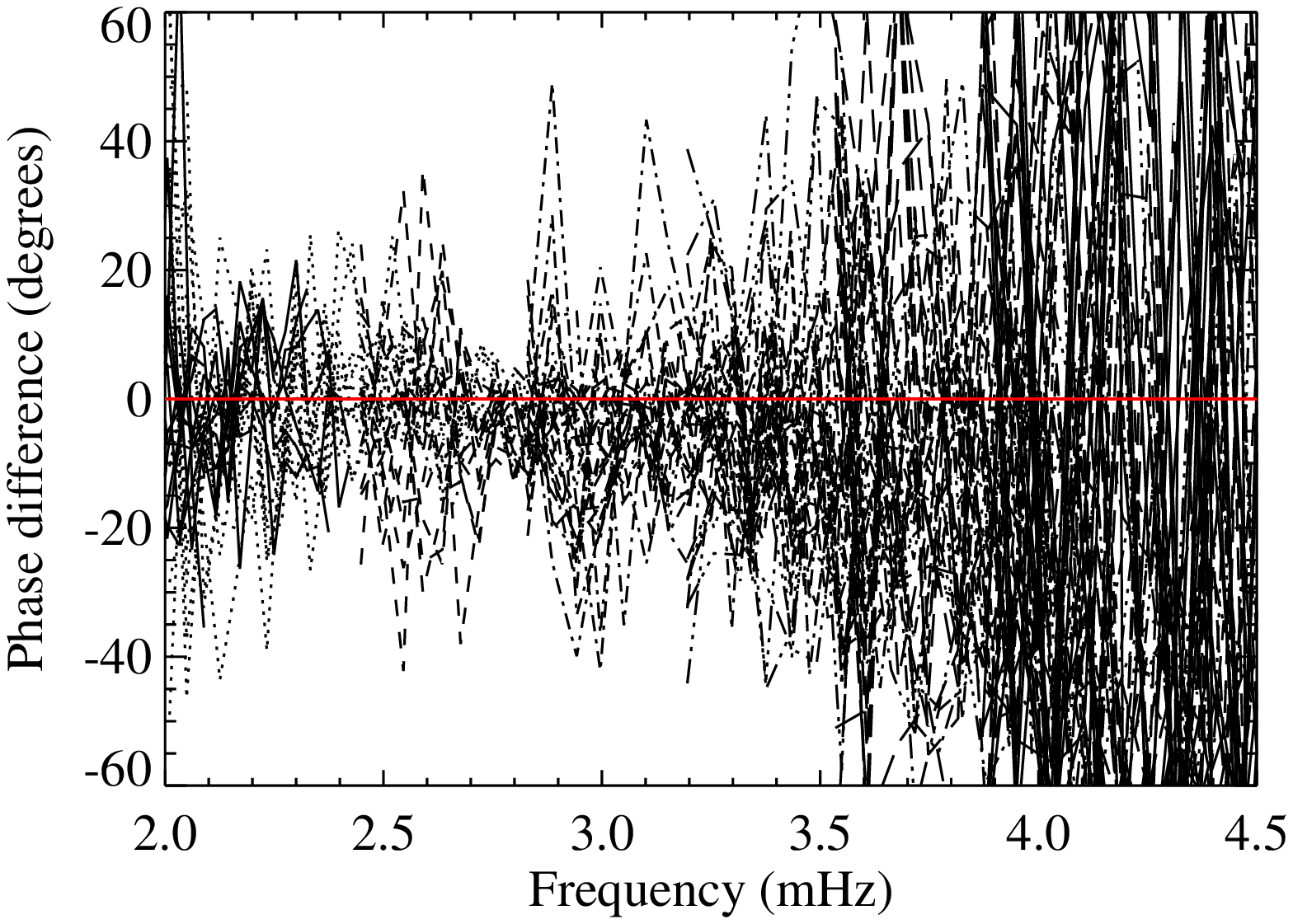}
\end{center}
\caption[]{
Phase difference $\phi_h$ between $20^\circ$ and $50^\circ$ from the center position.
The line styles are the same as in Fig.~\ref{fig:results0}.
}
\label{fig:phih}
\end{figure}

\begin{figure}
\begin{center}
\includegraphics[width=0.95\columnwidth]{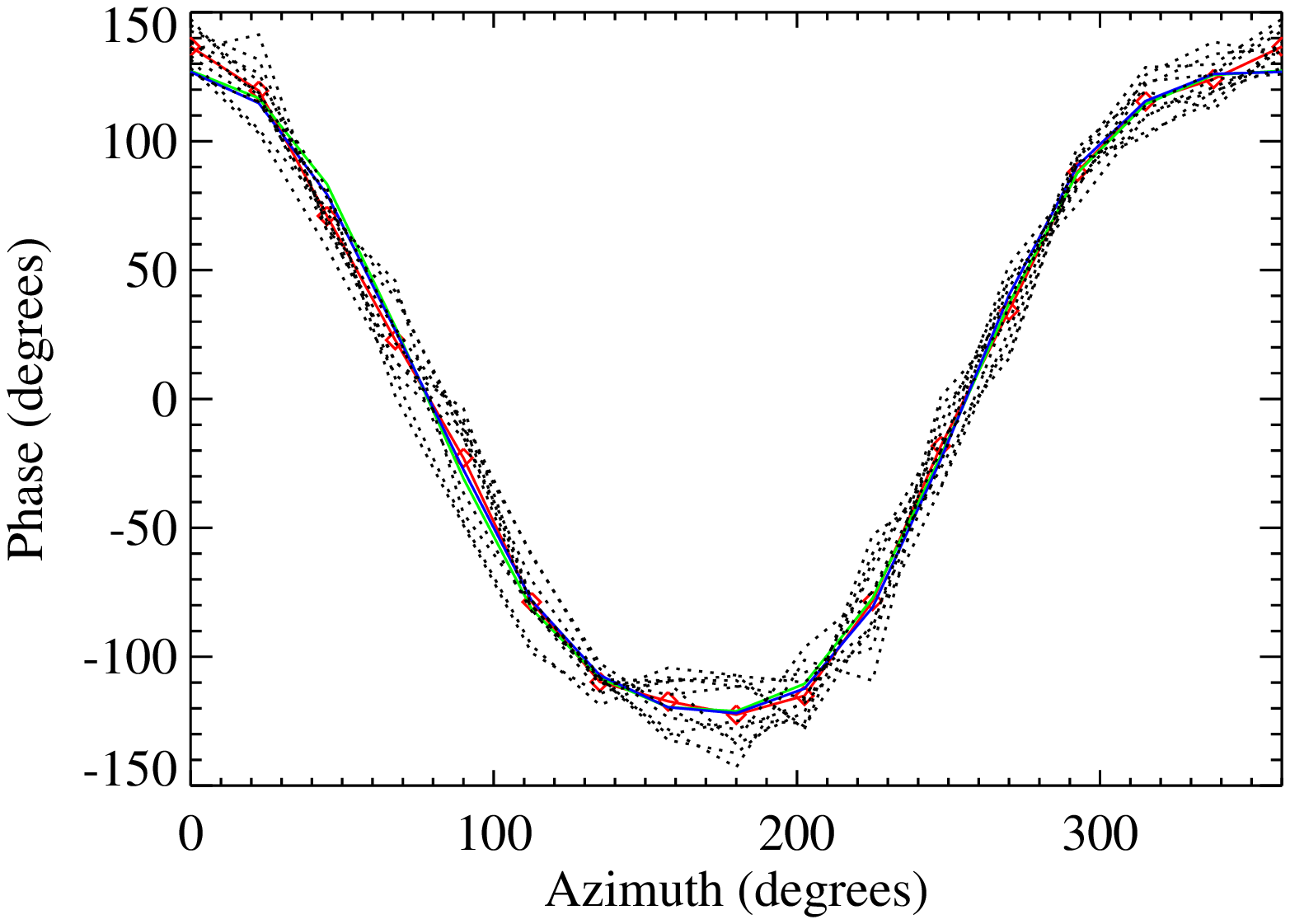}
\end{center}
\caption[]{
Phase for the f-mode and fits. Dotted lines show the modes with frequencies between roughly 2.033~mHz and 2.300~mHz. 
The red line and diamonds show the average. 
The green line shows the average of the fits over the frequency range, assuming $\phi_h=0$.
The blue line shows the fits with $\phi_h$ as a free parameter.
}
\label{fig:fitphih}
\end{figure}

Potentially $\phi_h$ could be nonzero due to, for example, nonadiabatic effects. To that end, fits were made using the full expression in Eq.~\ref{phip}. 
As illustrated in Fig.~\ref{fig:phip} the main effect of $\phi_h$ on the cross-spectrum phase is a skewness that only appears when $c_t$ is large and the azimuths are far from $180^\circ$. It would thus be expected that the best signal is in low $n$ modes when the azimuth difference is substantial.
Results from fits of such modes
are shown in Fig. \ref{fig:phih}, and an example of the improvement in the fit in Fig.~\ref{fig:fitphih}.
While the change in the fitted phase of the cross-spectrum is quite small (consistent with the minimal change in the fits), the scatter in the measured $\phi_h$ is small enough to see that $\phi_h$ is small in an absolute sense (small compared to a radian). On the other hand the scatter is almost certainly too large to conclude that the small negative bias is significant.

\section{Discussion}
\label{Discussion}
Unfortunately, it was not   possible to obtain the exact form of the deviation, as geometric problems got in the way. 
Having said that, the measured $c_t$ values agree quite well with those expected for adiabatic oscillations in a standard solar model when appropriate image shifts are chosen.
Nonetheless, it is worth   considering whether various observational or data analysis issues could affect the results significantly.

The short duration of the observations means that it is not possible to resolve the modes in the spectra. Rather, fitted values from a standard fit near the same time and positions were used.
To see why this is unlikely to present a problem we note that $c_t$ is the result of how a wave behaves near the surface and is not related to the existence of a resonance.
As expected, the results obtained by shifting the frequencies by half of a resolution element only deviate slightly from the reference results.

Another potential issue is that the difference in how the instruments sample the atmosphere, combined with the wave properties varying with height could change the results.
To investigate this, new Dopplergrams were calculated with different weights from those in
Eqs.~\ref{eq:cos} and \ref{eq:sin}, selected to better match the height at zero velocity. This did not cause a substantial change.

Using locations with different viewing angles for SO/PHI and HMI does change the results somewhat. This is likely due to having different observing heights for the two instruments combined with a variation of the wave properties with height.
However, changing the viewing angles also means that data closer to the limb is used. 
Especially for SO/PHI, which has a lower resolution on the Sun, this causes the foreshortening to be even more severe. For these reasons we did not use these results   here, but it may be worthwhile to revisit this issue when data with other viewing angles become available in the future.

Another  concern is   that the rather large tiles ($15^\circ \times 15^\circ$) might result in problems as the viewing geometry   changes across them. Similarly an error on the optical distortion, which locally    results in a shift,  with large enough tiles would result in a smearing for the SO/PHI and HMI datasets relative to each other. One more  concern is that the foreshortening would result in a variation in sensitivity across the tiles. To check for these effects the tiles were apodized to a $7.5^\circ$ diameter instead of the standard $15^\circ$ diameter. As expected, the main effect of this is a large increase in the noise. 
Beyond this there are no substantial changes.

While the inferred $c_t$ is almost perfectly orthogonal to the inferred offsets in the transverse direction and in time, it is nonetheless interesting to consider the origin of those shifts.
A large contribution to the transverse shift is undoubtedly geometrical errors. In particular residual distortion. The same also applies to the shift in the between-spacecraft direction, but here we do not know the amount as it is degenerate with $c_t$, as discussed earlier.

The time offset is more complicated. We recall that the time used to label the SO/PHI observations is the midpoint of the observation across the line. This is not necessarily the effective time of observation as the line is not symmetrically placed relative to the center tuning position. Furthermore, when the line is Doppler shifted (mainly due to the solar rotation) the sensitivity as a function of wavelength changes,
resulting in an effective time of observation that is spatially variable. A simple model of this effect indicates that it is comparable to the observed shifts.
For HMI this is not an issue (to lowest order) as all the data used to make a Dopplergram are first interpolated in time to the target time.
It is also possible that there is an error in the internal timing that is not accounted for in one of the spacecraft or instruments. 

On the theoretical side,
issues to consider include nonadiabatic effects
such as radiative damping or wave-convection interactions.
When considering these effects it is important to keep in mind that the phase difference between the horizontal and vertical components is very close to the expected $90^\circ$.

\section{Conclusion}
\label{Conclusion}
While the results obtained are intriguing, it is clear that more work is needed.
First of all more data would improve the situation substantially.
Not only was the length of the time series used here (7.5~h) very short by helioseismology standards, the data also only covered a limited range of viewing angles with a substantial foreshortening.
To this end, further observations are planned for early 2023 with a longer duration and smaller viewing angle.

Further theoretical studies are probably also warranted.
A straightforward approach might be to study the oscillations in a simulation by performing a radiative transfer calculation to simulate the observations by SO/PHI and HMI.
Based on the results of this, a better theoretical understanding can hopefully be obtained, which in turn may also lead to an improved understanding of the center-to-limb systematics in helioseismology.

Nonetheless, with the results agreeing substantially better with $\ctad$ than with $\ctref$,
analyses assuming that $\ctref$ is a good approximation should probably be updated to instead use the values from an eigenfunction calculation.

%
%

\begin{acknowledgements}
The authors would like to thank Aaron Birch for useful discussions
and Damien Fournier for assistance with the eigenfunctions.
Solar Orbiter is a space mission of international collaboration
between ESA and NASA, operated by ESA.  We are grateful to the ESA SOC
and MOC teams for their support.  The German contribution to SO/PHI is
funded by the BMWi through DLR and by MPG central funds.
The Spanish contribution is funded by AEI/MCIN/10.13039/501100011033/ (RTI2018-096886-C5, PID2021-125325OB-C5, PCI2022-135009-2) and ERDF “A way of making Europe”; “Center of Excellence Severo Ochoa” awards to IAA-CSIC (SEV-2017-0709, CEX2021-001131-S); and a Ramón y Cajal fellowship awarded to DOS.
The French contribution is funded by CNES.
The HMI data are courtesy of NASA/SDO and the HMI science team.
The data were processed at the German Data Center for SDO (GDC-SDO),
funded by the German Aerospace Center (DLR).
\end{acknowledgements}

%
%

\bibliographystyle{aa}
\bibliography{PHI22.bib}

\end{document}